\documentclass[prb,aps,twocolumn]{revtex4}
\usepackage{graphicx,amsmath,bm}
\usepackage{dcolumn}

\begin{document}

\title{Quantum Correlations of Ideal Bose and Fermi Gases in the Canonical Ensemble}

\author{Kazumasa Tsutsui}
\author{Takafumi Kita}
\affiliation{Department of Physics, Hokkaido University,
Sapporo 060-0810, Japan}
\date{\today}

\begin{abstract}
We derive an expression for the reduced density matrices of ideal Bose and Fermi gases in the canonical ensemble,
which corresponds to the Bloch--De Dominicis (or Wick's) theorem in the grand canonical ensemble for normal-ordered products of operators.
Using this expression, we study one- and two-body correlations of homogeneous ideal gases with $N$ particles.
The pair distribution function $g^{(2)}(r)$ of fermions clearly exhibits antibunching with $g^{(2)}(0)=0$ due to the Pauli exclusion principle at all temperatures,
whereas that of normal bosons shows bunching with $g^{(2)}(0)\approx 2$, corresponding to the Hanbury Brown--Twiss effect.
For bosons below the Bose--Einstein condensation temperature $T_0$, an off-diagonal long-range order develops in the one-particle density matrix
to reach $g^{(1)}(r)=1$ at $T=0$, and the pair correlation starts to decrease towards $g^{(2)}(r)\approx 1$ at $T=0$.
The results for $N\rightarrow \infty$ are seen to converge to those of the grand canonical ensemble obtained by assuming
the average $\langle\hat\psi({\bm r})\rangle$ of the field operator $\hat\psi({\bm r})$ below $T_0$.
This fact justifies the introduction of the ``anomalous'' average $\langle\hat\psi({\bm r})\rangle\neq 0$ below $T_0$ in the grand canonical ensemble 
as a mathematical means of removing unphysical particle-number fluctuations to reproduce the canonical results in the thermodynamic limit.
\end{abstract}


\maketitle

\section{\label{sec:level1}Introduction}

Systems of identical particles exhibit unique correlations even without interactions, 
which originate solely from permutation symmetry.\cite{KitaText}
They manifest themselves in the pair distribution function $g^{(2)}(r)$, i.e., the relative probability of finding a pair of particles separated by distance $r$.
Specifically, normal ideal bosons show enhancement from $g^{(2)}(\infty)=1$ to $g^{(2)}(0)=2$, signaling an effective attraction between particles, 
whereas that of equal-spin fermions is reduced to $g^{(2)}(0)= 0$ as predicted by the Pauli exclusion principle.
The two distinct correlations are also measurable as bunching and antibunching effects in the experimental detection of particles,
as first observed by Hanbury Brown and Twiss with photons back in 1956.\cite{HBT56,HBT57}
Furthermore, $g^{(2)}(0)$ below the Bose--Einstein condensation (BEC) temperature $T_0$ should exhibit a conspicuous reduction\cite{KitaText,Naraschewski99,Gomes05,Bosse11,Wright12}
that may be used to monitor the development of coherence 
in condensates.\cite{KM97}

More generally, much attention has been focused on the $n$-body correlations $g^{(n)}(r)$ of quantum gases both
experimentally\cite{HBT56,Yasuda96,Burt97,Bloch00,Ottl05,Schellekens05,Rom06,Jeltes07,Haller11,Dall13,Schmitt14} 
and theoretically\cite{HBT57,Glauber63,KM97,Naraschewski99,Gomes05,Bosse11,Wright12,Penrose51,PO56,Yang} over many decades.
Among them, the one-body correlation $g^{(1)}(r)$ was used by Penrose\cite{Penrose51} and by Penrose and Onsager\cite{PO56}
to define BEC by
$g^{(1)}(\infty)>0$, i.e., the expectation called {\it the off-diagonal long-range order} by Yang.\cite{Yang}

However, there is a fundamental issue on the theoretical side that almost all the calculations 
have been performed in the grand canonical ensemble,\cite{HBT57,Naraschewski99,Gomes05,Bosse11}  
which is known to suffer from unphysically huge particle-number fluctuations below $T_0$.\cite{Fierz56,FHW70,Johnston70,ZUK77,Politzer96,WW97,HKK98}
Ad hoc approximations have been introduced to remove them, such as the introduction of the ``anomalous'' average $\langle\hat\psi({\bm r})\rangle$ 
of the field operator $\hat\psi({\bm r})$, which is apparently incompatible with the particle-number conservation.\cite{KitaText,Naraschewski99,Gomes05,Bosse11}
A calculation within the canonical ensemble was performed in this context, but with an approximation by discarding several terms.\cite{Wright12}
Hence, it is desirable to calculate one- and two-body correlations completely within the canonical ensemble, 
observe the approach to the thermodynamic limit, and check the validity of the approximations adopted in the grand canonical calculations.
We carry out such a calculation using a new formula for the reduced density matrices in the canonical ensemble,
which is the equivalent of the Bloch--De Dominicis (or Wick's) theorem in the grand canonical ensemble\cite{KitaText,BdD59,Gaudin60,Wick50} for normal-ordered products of operators.
Thus, the formula may also be used in perturbative calculations with respect to interactions in the canonical ensemble,
such as self-consistent Hartree-Fock calculations at finite temperatures with a fixed number of particles.

In Sect.\ \ref{sec:level2} we derive an analytic expression for the reduced density matrices of ideal gases in the canonical ensemble.
Section \ref{sec:level3} presents numerical results on the one- and two-body correlation functions of homogeneous Bose and Fermi gases in the canonical ensemble
 for different particle numbers at several temperatures in comparison with those of the grand canonical ensemble.
Concluding remarks are given in Sect.\ \ref{sec:level4}.
The Appendix proves that a superposition of the obtained formula for the reduced density matrices in terms of the particle number  reproduces 
 the Bloch--De Dominicis theorem in the grand canonical ensemble for normal-ordered products of operators.

\section{\label{sec:level2} Formula for Reduced Density Matrices}

\subsection{Definitions and main results}

We consider a canonical ensemble of $N$ identical particles with spin $s$ described by Hamiltonian $\hat{H}$.
The reduced $n$-particle density matrix ($n=1,2,\cdots,N$) is defined by\cite{KitaText}
\begin{align}
&\,\rho^{(n)}_{N}(\xi_1,\cdots,\xi_n;\xi'_1,\cdots,\xi_n')
\notag \\
\equiv &\,
\frac{1}{Z_N}{\rm Tr}\, e^{-\beta\hat H}\hat\psi^{\dagger}(\xi'_1)\cdots\hat\psi^{\dagger}(\xi'_n)\hat\psi(\xi_n)\cdots\hat\psi(\xi_1).
\label{rho^(n)-def}
\end{align}
Here $Z_N$ is the partition function with $Z_{0}\equiv 1$ by definition,
Tr denotes the trace, $\beta\equiv 1/k_{\rm B}T$ with $k_{\rm B}$ and $T$ the Boltzmann constant and temperature, respectively,
and $\hat\psi(\xi)$ is the field operator of bosons or fermions
with $\xi\equiv({\bm r},\alpha)$ denoting space (${\bm r}$) and spin ($\alpha=s,s-1,\cdots,-s$) coordinates.

The purpose of this section is to prove that Eq.\ (\ref{rho^(n)-def}) for {\em non-interacting systems} is expressible as
\begin{align}
&\,\rho^{(n)}_{N}(\xi_1,\cdots,\xi_n;\xi'_1,\cdots,\xi_n')
\notag \\
=&\, \frac{1}{Z_N}\sum_{m=n}^N \sigma^{m-n} Z_{N-m} \sum_{\ell_1= 1}^m\cdots\sum_{\ell_{n}=1}^m\delta_{\ell_1+\cdots+\ell_{n},m}
\notag \\
&\,\times
\sum_{\hat P_{n}}\sigma^{P_{n}} \prod_{j=1}^n \phi_{\ell_{j}}(\xi_j,\xi_{p_j}').
\label{rho^(n)}
\end{align}
Here $\sigma=1$ and $-1$ for bosons and fermions, 
$\hat P_n$ denotes a permutation with $n$ elements,\cite{KitaText}
\begin{align}
\hat P_n\equiv\left(\begin{array}{cccc}1&2&\cdots &n\\ p_1&p_2&\cdots&p_n\end{array}\right),
\label{P_n}
\end{align}
$\sigma^{P_{n}}=1$ and $\sigma$ for even and odd permutations, respectively,
and $\phi_\ell (\xi,\xi')$ is defined in terms of the one-particle eigenvalues 
$\varepsilon_{q}$ and orthonormal eigenfunctions $\varphi_q(\xi)=\langle\xi|q\rangle$ as
\begin{equation}
\phi_\ell (\xi,\xi')\equiv\sum_{q}e^{-\ell\beta\varepsilon_{q}}\langle \xi|q\rangle\langle q|\xi'\rangle,
\label{phi_n}
\end{equation}
where $q$ distinguishes one-particle eigenstates.

Let us write Eq.\ (\ref{rho^(n)}) for $n=1,2$ explicitly for later purposes:
\begin{subequations}
\label{rho12}
\begin{align}
\rho_N^{(1)}(\xi_1,\xi_1')
=\frac{1}{Z_N}\sum_{m=1}^{N}\sigma^{m-1}Z_{N-m} \phi_m(\xi_1,\xi_1'),
\label{rho1}
\end{align}
\begin{align}
&\,\hspace{-5mm}\rho^{(2)}_N(\xi_1,\xi_2;\xi_1',\xi_2')
\notag \\
=&\,\frac{1}{Z_N}\sum_{m=2}^{N}\sigma^{m} Z_{N-m} \sum_{\ell=1}^{m-1}\bigl[\phi_\ell(\xi_1,\xi_1')\phi_{m-\ell}(\xi_2,\xi_2')
\notag \\
&\, +\sigma \phi_\ell(\xi_1,\xi_2')\phi_{m-\ell}(\xi_2,\xi_1')\bigr].
\label{rho2}
\end{align}
\end{subequations}
Note that $\rho^{(2)}_N$ is not expressible concisely in terms of $\rho_N^{(1)}$,
unlike the case of the grand canonical ensemble.

Two comments on Eq.\ (\ref{rho^(n)}) are in order.
First, Eq.\ (\ref{rho1}) enables us to reproduce
the following recurrence relation for the partition function:\cite{Landsberg61}
\begin{align}
Z_N=\frac{1}{N}\sum_{m=1}^N \sigma^{m-1} Z_{N-m} S_m,\hspace{5mm}S_m\equiv \sum_q e^{-m\beta\varepsilon_q} ,
\label{Z_N-rec}
\end{align}
by setting $\xi_1'=\xi_1$, subsequently integrating over $\xi_1$, and using the sum rule:\cite{KitaText}
\begin{align}
\int \rho_N^{(1)}(\xi_1,\xi_1)d\xi_1=N ,
\label{rho1-sum}
\end{align}
where the integration over $\xi_1$ involves summation over $\alpha_1=s,s-1,\cdots,-s$ and integration over ${\bm r}_1$.
Second, Eq.\ (\ref{rho^(n)}) corresponds to the Bloch--De Dominicis (or Wick's) theorem in the grand canonical ensemble
for normal-ordered products of operators.\cite{KitaText,BdD59,Gaudin60,Wick50}
Indeed, multiplying Eq.\ (\ref{rho^(n)}) by $Z_N e^{\beta\mu N}$, with $\mu$ the chemical potential, and 
summing over $N$, we obtain the $n$-particle density matrix $\rho_{\rm G}^{(n)}$ in the grand
canonical ensemble as
\begin{align}
&\,\rho_{\rm G}^{(n)}(\xi_1,\cdots,\xi_{n};\xi_1',\cdots,\xi_{n}')=\sum_{\hat P_{n}}\sigma^{P_{n}}\prod_{j=1}^n \langle \hat\psi^\dagger(\xi_j')\hat\psi(\xi_{p_j})\rangle ,
\label{Wick-GC}
\end{align}
where $\langle\hat A\rangle\!\equiv\! Z_{\rm G}^{-1}{\rm Tr}e^{-\beta(\hat H-\mu \hat N)}\hat A$ with $Z_{\rm G}\!\equiv\! {\rm Tr}e^{-\beta(\hat H-\mu \hat N)}$ the grand partition function;
see the Appendix for details of the derivation.
This is exactly the Wick decomposition for the average $\langle\hat\psi^{\dagger}(\xi'_1)\cdots\hat\psi^{\dagger}(\xi'_n)\hat\psi(\xi_n)\cdots\hat\psi(\xi_1)\rangle$
without the expectation $\langle\psi(\xi)\rangle$, 
expressed solely in terms of the one-particle density matrix $\rho_{\rm G}^{(1)}(\xi_1,\xi_{1}')=\langle\hat\psi^\dagger(\xi_1')\hat\psi(\xi_{1})\rangle$. 

\subsection{Preliminaries for proof}

To prove Eq.\ (\ref{rho^(n)}),
we expand the field operators as
\begin{align}
\hat\psi(\xi)=\sum_{q}\langle \xi|q\rangle \hat c^{}_{q},\hspace{5mm} \hat\psi^\dagger(\xi)=\sum_{q} \hat c^\dagger_{q}\langle q|\xi\rangle.
\label{fo}
\end{align}
Our Hamiltonian is given explicitly in terms of $\hat c_{q}$, $\hat c^\dagger_{q}$, and $\varepsilon_q$ as\cite{KitaText}
\begin{align}
\hat H=\sum_{q}\varepsilon_{q} \hat c^{\dagger}_{q}\hat c^{}_{q},
\label{H0}
\end{align}
and the eigenstates of $\hat H$ are expressible as
\begin{align}
|\nu\rangle\equiv|n_{q_1}n_{q_2}\cdots n_{q_M}\rangle=\frac{(\hat c^{\dagger}_{q_1})^{n_{q_1}}}{\sqrt{n_{q_1}!}}\cdots\frac{(\hat c^{\dagger}_{q_M})^{n_{q_M}}}{\sqrt{n_{q_M}!}}|0\rangle,
\end{align}
where $n_{q_j}=1,2,3,\cdots$ ($n_{q_j}=1$) for bosons (fermions) in the canonical ensemble with $n_{q_1}+n_{q_2}+\cdots +n_{q_M}=N$.
The corresponding eigenvalues are given by
\begin{align}
E_\nu=\sum_{j=1}^M n_{q_j} \varepsilon_{q_j} .
\label{E_nu}
\end{align}
The partition function of ideal gases can be written in two different forms as
\begin{align}\label{ZN}
Z_N=&\,\sum_{\nu}e^{-\beta E_\nu}
\notag \\
=&\,\sum_{q_1,\cdots,q_N}w_{N}(q_1,\cdots,q_N)e^{-\beta(\varepsilon_{q_1}+\cdots+\varepsilon_{q_N})}.
\end{align}
The first sum is over all possible distinct sets of occupation numbers, 
whereas the second one can be performed independently over each $q_j$ ($j=1,2,\cdots,N$) 
by using a factor that obeys the following recurrence relation:\cite{Borrmann}
\begin{align}\label{w_q}
&\,w_N(q_1,\cdots,q_N)
\notag \\
=&\,\frac{1}{N}w_{N-1}(q_1,\cdots,q_{N-1}) \left(1+\sigma\sum_{j=1}^{N-1}\delta_{q_jq_N}\right),
\end{align}
with $w_1(q_1)=1$. Note that $1/w_N$ for $\sigma=1$ is the number of ways of arranging the set $(q_1,\cdots,q_N)$, which may contain identical states.
On the other hand, $w_N$ for $\sigma=-1$ has the effect of removing multiple occupancies from any one-particle state.
In both cases, we can express $w_N$ for every distinct permutation of the state
\begin{subequations}
\label{w_N-2}
\begin{align}
\nu=(\,\underbrace{q_{1},\cdots,q_{1}}_{n_{1}},\underbrace{q_{2},\cdots,q_{2}}_{n_{2}},\cdots\cdots,\underbrace{q_M,\cdots,q_M}_{n_M}\,) 
\end{align}
as
\begin{align}
w_N(\nu)=\frac{n_1!\cdots n_M!}{N!} ,
\end{align}
\end{subequations}
with $n_1=\cdots =n_M=1$ and $M=N$ for fermions.

\subsection{Proof of Eq.\ (\ref{rho^(n)})}

We prove Eq.\ (\ref{rho^(n)}) by induction.
First, the case for $n=N$ is shown to hold
by using the orthonormal eigenfunctions\cite{KitaText} $\Phi_\nu(\xi_1,\cdots,\xi_N)\equiv\langle\xi_1\cdots\xi_N|n_{q_1}\cdots n_{q_M}\rangle$ as
\begin{align}
&\,\rho_N^{(N)}(\xi_1,\cdots,\xi_N;\xi_1',\cdots,\xi_N')
\notag \\
\equiv&\,\frac{N!}{Z_N}\sum_{\nu} e^{-\beta E_\nu}\Phi_\nu(\xi_1,\cdots,\xi_N)\Phi_\nu^*(\xi_1',\cdots,\xi_N') 
\notag \\
= &\,\frac{N!}{Z_N}\sum_{\nu}e^{-\beta E_\nu}\sum_{\hat P_N'\hat P_N}
\frac{\sigma^{P_N'+P_N}}{N!n_1!\cdots n_M !}\prod_{j=1}^N \langle \xi_j|q_{p_j'}\rangle \langle q_{p_j} | \xi_j'\rangle
\notag \\
&\, \hspace{5mm}\mbox{$\leftarrow$ multiply the contribution of $\hat P_N'=1$ by $N!$}
\notag \\
= &\,\frac{1}{Z_N}\sum_{\nu}e^{-\beta E_\nu}\frac{N!}{n_1!\cdots n_M !}
\sum_{\hat P_N}
\sigma^{P_N}\prod_{j=1}^N \langle \xi_j|q_j\rangle \langle q_{p_j} | \xi_j'\rangle
\notag \\
&\, \hspace{5mm}\mbox{$\leftarrow$ the same transformation as Eq.\ (\ref{ZN}) }
\notag \\
&\, \hspace{10mm}\mbox{using Eq.\ (\ref{w_N-2})}
\notag \\
= &\,\frac{1}{Z_N}\sum_{q_1,\cdots,q_N}e^{-\beta(\varepsilon_{q_1}+\cdots +\varepsilon_{q_N})} \sum_{\hat P_N}
\sigma^{P_N} \prod_{j=1}^N \langle \xi_j|q_j\rangle \langle q_{p_j} | \xi_j'\rangle
\notag \\
= &\,\frac{1}{Z_N} \sum_{\hat P_N}\sigma^{P_N}
\prod_{j=1}^N \sum_{q_j}e^{-\beta\varepsilon_{q_j}} \langle \xi_j|q_j\rangle \langle q_j | \xi_{p_j}'\rangle
\notag \\
= &\,\frac{1}{Z_N} \sum_{\hat P_N}\sigma^{P_N}\prod_{j=1}^N
\phi_1(\xi_{j},\xi_{p_j}').
\label{rho^(N)-proof}
\end{align}
The last expression is  identical to Eq.\ (\ref{rho^(n)}) for $n=N$.

Now, suppose that Eq.\ (\ref{rho^(n)}) holds true for a certain $n$. 
Then $\rho_N^{(n-1)}(\xi_1,\cdots,\xi_{n-1};\xi_1',\cdots,\xi_{n-1}')$ can be obtained 
by setting $\xi_n'=\xi_n$ in Eq.\ (\ref{rho^(n)}) and integrating over $\xi_n$ as follows:\cite{KitaText}
\begin{subequations}
\label{rho^(N)-proof123}
\begin{align}
&\,\rho_N^{(n-1)}(\xi_1,\cdots,\xi_{n-1};\xi_1',\cdots,\xi_{n-1}')
\notag \\
\equiv &\,\frac{1}{N\!-\!n\!+\!1} \int d\xi_n\, \rho_N^{(n)}(\xi_1,\cdots,\xi_{n-1},\xi_n;\xi_1',\cdots,\xi_{n-1}',\xi_n)
\notag \\
=&\,\sum_{m=n}^N   \frac{\sigma^{m-n} Z_{N-m}}{(N-n+1)Z_N}\sum_{\ell_1= 1}^m\cdots\sum_{\ell_{n}=1}^m\delta_{\ell_1+\cdots+\ell_{n},m}
\sum_{\hat P_{n}}\sigma^{P_{n}}
\notag \\
&\,\times
 \prod_{j=1}^{n-1}
\sum_{q_j} e^{-\ell_j\beta\varepsilon_{q_j}} \langle\xi_j|q_j\rangle \langle q_{p_j}|\xi_j'\rangle
\sum_{q_n} \delta_{q_nq_{p_n}}e^{-\ell_n\beta\varepsilon_{q_n}} .
\label{rho^(N)-proof1}
\end{align}
Let us consider the cases with $p_n=n$ and those with $p_n\leq n-1$ separately in treating the last sum.
Permutations for $p_n=n$ are expressible as
\begin{align}
\hat P_n=\left(\begin{array}{cccc}1&\cdots &n-1&n\\ p_1&\cdots&p_{n-1}& n \end{array}\right)=
\left(\begin{array}{c}n\\ n \end{array}\right)\hat P_{n-1} ,
\notag
\end{align}
whereas those for $p_n=i\leq n-1$ can be transformed as\cite{KitaText}
\begin{align}
\hat P_n=\left(\begin{array}{cccc}1&\cdots &n-1&n\\ p_1&\cdots&p_{n-1}& i \end{array}\right)=
\left(\begin{array}{cc}i & n\\ n & i \end{array}\right)\hat P_{n-1} ',
\notag
\end{align}
where $\hat P_{n-1}'$ is some permutation with $n-1$ elements.
This fact enables us to replace the sum over $\hat P_n$ in Eq.\ (\ref{rho^(N)-proof1}) by that over $\hat P_{n-1}$ 
with the introduction of factors $1$ and $\sigma$ for $p_n=n$ and $p_n\leq n-1$, respectively.
We thereby obtain
\begin{align}
&\,\rho_N^{(n-1)}(\xi_1,\cdots,\xi_{n-1};\xi_1',\cdots,\xi_{n-1}')
\notag \\
=&\, \sum_{m=n}^N  \frac{\sigma^{m-n}  Z_{N-m}}{(N\!-\!n\!+\!1)Z_N}\sum_{\ell_1= 1}^m\cdots\sum_{\ell_{n}=1}^m\delta_{\ell_1+\cdots+\ell_{n},m}
\sum_{\hat P_{n-1}}\sigma^{P_{n-1}}
\notag \\
&\,\times 
 \prod_{j=1}^{n-1} \sum_{q_j} e^{-\ell_j\beta\varepsilon_{q_j}} \langle\xi_j|q_j\rangle \langle q_{j}|\xi_{p_j}'\rangle
\notag \\
&\,\times 
\left(  S_{\ell_n}
+\sigma \sum_{i=1}^{n-1}e^{-\ell_n\beta\varepsilon_{q_i}}\right) .
\label{rho^(N)-proof2}
\end{align}
The first contribution in the round bracket can be transformed further by using $m'\equiv m-\ell_n$
as follows:
\begin{align}
&\, \sum_{m=n}^N \sigma^{m-n}Z_{N-m} \sum_{\ell_n= 1}^m \delta_{\ell_1+\cdots+\ell_{n},m}S_{\ell_n} 
\notag \\
=&\, 
\sum_{m'=n-1}^N \sum_{\ell_n=1}^{N-m'}  \sigma^{m'+\ell_n-n}Z_{N-m'-\ell_n} S_{\ell_n} \delta_{\ell_1+\cdots+\ell_{n-1},m'}
\notag \\
=&\, 
\sum_{m'=n-1}^N   \sigma^{m'-n+1}(N-m')Z_{N-m'} \delta_{\ell_1+\cdots+\ell_{n-1},m'} ,
\notag 
\end{align}
where we used Eq.\ (\ref{Z_N-rec}).
For the second contribution in the round brackets of Eq.\ (\ref{rho^(N)-proof1}), the sums over $(l_i,l_n)$
can be transformed with a change of variables $\ell_i'\equiv \ell_i+\ell_n$ as
\begin{align}
&\,\sum_{\ell_i=1}^m \sum_{\ell_n= 1}^m{\rm e}^{-(\ell_i+\ell_n)\beta \varepsilon_{q_i}}\delta_{\ell_1+\cdots+\ell_{n},m} 
\notag \\
= &\,  \sum_{\ell_i'= 2}^m\sum_{\ell_n= 1}^{\ell_i'-1} {\rm e}^{-\ell_i'\beta \varepsilon_{q_i}}\delta_{\ell_1+\cdots+\ell_i'+\cdots+\ell_{n-1},m} 
\hspace{10mm}(\ell_i'\rightarrow \ell_i)
\notag \\
=&\,\sum_{\ell_i= 2}^m(\ell_i-1){\rm e}^{-\ell_i\beta \varepsilon_{q_i}}\delta_{\ell_1+\cdots+\ell_{n-1},m} 
\notag \\
=&\,\sum_{\ell_i=1}^m(\ell_i-1){\rm e}^{-\ell_i\beta \varepsilon_{q_i}}\delta_{\ell_1+\cdots+\ell_{n-1},m} .
\notag
\end{align}
The subsequent summation of $\ell_i-1$ over $i=1,\cdots,n-1$ with the constraint $\delta_{\ell_1+\cdots+\ell_{n-1},m}$ yields $m-(n-1)$.
Substituting the above two results into Eq.\ (\ref{rho^(N)-proof2}), we obtain
\begin{align}
&\,\rho_N^{(n-1)}(\xi_1,\cdots,\xi_{n-1};\xi_1',\cdots,\xi_{n-1}')
\notag \\
=&\, \sum_{m=n-1}^N \frac{\sigma^{n-1+m} Z_{N-m}}{(N-n+1)Z_N} [(N-m)+(m-n+1)] 
\notag \\
&\,\times 
\sum_{\ell_1= 1}^m\cdots\sum_{\ell_{n-1}=1}^m\delta_{\ell_1+\cdots+\ell_{n-1},m}
\sum_{\hat P_{n-1}}\sigma^{P_{n-1}}
\notag \\
&\,\times
 \prod_{j=1}^{n-1} \phi_{\ell_j}(\xi_j,\xi_{p_j}')  ,
\label{rho^(N)-proof3}
\end{align}
\end{subequations}
which is exactly Eq.\ (\ref{rho^(n)}) with the replacement $n\rightarrow n-1$.
This completes our proof of Eq.\ (\ref{rho^(n)}).

\section{\label{sec:level3}Correlation Functions of Homogeneous Systems}

\subsection{Correlation functions}

We study one- and two-body correlations of
homogeneous ideal gases in the canonical ensemble based on Eq.\ (\ref{rho12}).
Specifically, we consider identical particles with mass $m$ in a box of volume $V=L^3$ with periodic boundary conditions.
One-particle eigenstates are distinguished in terms of the wave vector ${\bm k}=(2\pi n_x/L,2\pi n_y/L,2\pi n_z/L)$ by $q=({\bm k},\alpha)$, 
whose eigenvalues and eigenfunctions are given by
\begin{equation} 
\varepsilon_{k}=\frac{\hbar^2 k^2}{2m},\hspace{10mm}\varphi_{{\bm k}\alpha}(\xi_1)=\frac{1}{\sqrt{V}}e^{i{\bm k}\cdot{\bm r}_1} \delta_{\alpha\alpha_1} .
\end{equation}
Equation (\ref{phi_n}) can be written as
\begin{align}
\phi_\ell (\xi_1,\xi_2)=&\,\frac{\delta_{\alpha_1\alpha_2}}{V}\sum_{\bm k}e^{-\ell\beta\varepsilon_{k}+i{\bm k}\cdot({\bm r}_1-{\bm r}_2)}
\notag \\
\equiv&\, \delta_{\alpha_1\alpha_2}f_\ell({\bm r}_1-{\bm r}_2).
\label{QQ1}
\end{align}
Note that $\phi_\ell (\xi,\xi)=f_\ell({\bf 0})=S_\ell/(2s+1)V$ as can be seen from Eq.\ (\ref{Z_N-rec}). 
Thus, the diagonal element $\phi_\ell (\xi,\xi)$ does not depend on $\xi$ at all  for homogeneous systems.
Let us introduce one- and two-body correlation functions of homogeneous systems as
\begin{subequations}
\label{g^(12)-def}
\begin{align}
g^{(1)}_{\alpha_1\alpha_2}(|{\bm r}_1-{\bm r}_2|)\equiv&\,\frac{(2s+1)V}{N}\rho^{(1)}_{N}(\xi_1,\xi_2),
\label{g^(1)-def}
\\
g^{(2)}_{\alpha_1\alpha_2}(|{\bm r}_1-{\bm r}_2|)\equiv&\,\left[\frac{(2s+1)V}{N}\right]^2 \rho^{(2)}_{N}(\xi_1,\xi_2;\xi_1,\xi_2).
\label{g^(2)-def}
\end{align}
\end{subequations}
Equation (\ref{g^(2)-def}) is also called the pair distribution function.
We set $\xi_2=\xi_1$ in Eq.\ (\ref{g^(1)-def}), integrate the resulting expression over $\xi_1=({\bm r}_1,\alpha_1)$,
and use Eq.\ (\ref{rho1-sum}). We then find that
\begin{align}
\sum_{\alpha_1}g^{(1)}_{\alpha_1\alpha_1}(0)=2s+1.
\notag
\end{align}
We also integrate Eq.\ (\ref{g^(2)-def}) over $(\xi_1,\xi_2)$ and use the first relation of Eq.\ (\ref{rho^(N)-proof1}) for $n=1,2$.
We thereby obtain 
\begin{align}
\frac{N}{(2s+1)^2 V}\sum_{\alpha_1\alpha_2}\int d^3 r \left[ 1-g^{(2)}_{\alpha_1\alpha_2}(r)\right] =1.
\notag
\end{align}
Noting that (i) $g^{(n)}_{\alpha_1\alpha_1}(r)$ does not depend on $\alpha_1$ in the absence of magnetic fields
 and (ii) $g^{(2)}_{\alpha_1\alpha_2}(r)=1$ for $\alpha_1\neq \alpha_2$,
we can express the above results in terms of the diagonal element $g^{(n)}(r) \equiv g^{(n)}_{\alpha\alpha}(r)$ as
\begin{subequations}
\label{g12-sum}
\begin{align}
g^{(1)}(0)=1,
\label{g^(1)-sum}
\end{align}
\begin{align}
\frac{N}{(2s+1) V}\int d^3 r \left[1- g^{(2)}(r)\right] =1.
\label{g^(2)-sum}
\end{align}
\end{subequations}

We will compare the functions in Eq.\ (\ref{g^(12)-def}) with the following ones in the grand canonical ensemble:\cite{KitaText}
\begin{subequations}
\label{g^(12G)}
\begin{align}
g^{(1{\rm G})}_{\alpha_1\alpha_2}(r)\equiv \frac{N_0}{N}+\delta_{\alpha_1\alpha_2}\ell(k_{\rm Q}|{\bm r}_1-{\bm r}_2|) ,
\label{g^(1G)}
\end{align}
\begin{equation}
g^{(2{\rm G})}_{\alpha _{1}\alpha _{2}}(r)= 1+\delta_{\alpha_1\alpha_2}\sigma
\left\{[\ell (k_{\rm Q}r)]^2+2\frac{N_0}{N}\ell (k_{\rm Q}r)\right\} ,
\label{g^(2G)}
\end{equation}
\end{subequations}
which are applicable to both bosons ($\sigma=1$) and fermions ($\sigma=-1$, $N_0=0$) 
and obtained for bosons below $T_0$ by assuming the anomalous average $\langle \hat\psi(\xi)\rangle$.
Here $N_0$ is the number of condensed particles, which becomes finite for bosons below $T_0$ and is given by $N_0=N[1-(T/T_0)^{2/3}]$,\cite{KitaText}
$k_{\rm Q}$ is defined together with $T_{\rm Q}$ by
\begin{align}
k_{\rm Q}\equiv\frac{\pi}{L}\left(\frac{N}{2s+1}\right)^{1/3} ,\hspace{10mm}T_{\rm Q}\equiv\frac{\hbar^2k_{\rm Q}^2}{2mk_{\rm B}},
\label{Qunit}
\end{align}
and $\ell(x)$ denotes
\begin{equation}
\ell(x)\equiv\frac{\pi}{4}\int_0^\infty\frac{1}{e^{(\tilde\epsilon-\tilde{\mu})/\tilde{T}}-\sigma}\frac{\sin(\tilde{\epsilon}^{1/2}x)}{x}d\tilde{\epsilon},
\end{equation}
with $\tilde\mu\equiv \mu/k_{\rm B}T_{\rm Q}$ and $\tilde T\equiv T/T_{\rm Q}$.
Note that $N$ in the grand canonical ensemble denotes $\langle N\rangle$.
The BEC temperature $T_0$ for bosons and the Fermi temperature $T_{\rm F}$ for fermions are given in terms of $T_{\rm Q}$
by\cite{KitaText}
\begin{align}
T_0=0.671T_{\rm Q},\hspace{10mm} T_{\rm F}=1.54T_{\rm Q}.
\end{align}
For the sum rules in the grand canonical ensemble, 
Eq.\ (\ref{g^(1G)}) also satisfies Eq.\ (\ref{g^(1)-sum}), whereas Eq.\ (\ref{g^(2G)}) obeys
\begin{align}
\frac{\langle N\rangle}{(2s+1)V}\int d^3 r\left[1-g^{(2{\rm G})}(r)\right]=1-\frac{\langle N^2\rangle-\langle N\rangle^2}{\langle N\rangle},
\label{g^(2G)-sum}
\end{align}
where $\langle\cdots\rangle$ denotes the grand canonical average; Eq.\ (\ref{g^(2G)-sum}) can be shown similarly as Eq.\ (\ref{g^(2)-sum}).
One of our main interests is how $g^{(2{\rm G})}(r)$ is different from $g^{(2)}(r)$ bringing about the additional (fluctuation) term in Eq.\ (\ref{g^(2G)-sum}).

\subsection{Numerical procedures}

Equation (\ref{g^(12)-def}) can be evaluated numerically from Eqs.\ (\ref{rho12}) and (\ref{Z_N-rec}).
The key quantity here is the function $f_\ell({\bm r})$ defined in Eq.\ (\ref{QQ1}).
It is expressible as
\begin{subequations}
\label{Poisson}
\begin{align}
f_\ell({\bm r})=\frac{1}{L^3}\prod_{\lambda=x,y,z}\sum_{n_\lambda=-\infty}^\infty \exp \left[-\frac{4\ell n_\lambda^2}{\tau}+2in_\lambda\rho_\lambda\right] ,
\label{Poisson1}
\end{align}
with $\tau\!\equiv\! [N/(2s+1)]^{2/3}T/T_{\rm Q}$ and ${\bm \rho}\!\equiv\! [(2s+1)/N]^{1/3}k_{\rm Q}{\bm r}$.
Thus, $f_\ell$ is periodic in $\rho_\lambda$ with period $\pi$.
Using the Poisson summation formula, we can also transform Eq.\ (\ref{Poisson1}) into
\begin{align}
&\, f_\ell({\bm r})
\notag \\
=&\,\frac{1}{L^3}\left(\frac{\pi\tau}{4\ell}\right)^{3/2}\prod_{\lambda=x,y,z}\sum_{n_\lambda=-\infty}^\infty \exp \left[-\frac{\pi^2\tau}{4\ell}
\left(n_\lambda-\frac{\rho_\lambda}{\pi}\right)^2\right] .
\label{Poisson2}
\end{align}
\end{subequations}
Since it depends only on $r\equiv |{\bm r}|$, we can obtain $f_\ell({\bm r})$ by choosing ${\bm r}$ along the $x$-axis, for example.
Equations (\ref{Poisson1}) and (\ref{Poisson2}) were found to give good convergence in the summation over $n_\lambda$ for $\tau/4\ell\lesssim0.5$ and $\tau/4\ell\gtrsim0.5$, respectively.
Round-off errors that may accumulate significantly in evaluating Eqs.\ (\ref{rho12}) and (\ref{Z_N-rec}) were removed by high-precision calculations with Mathematica. 
We confirmed that the sum rule in Eq.\ (\ref{g12-sum}) is satisfied for each of our numerical calculations for $N=50,100,1000$.

\subsection{Numerical results}

\begin{figure}[t]
\begin{center}
  \includegraphics[width=0.8\linewidth]{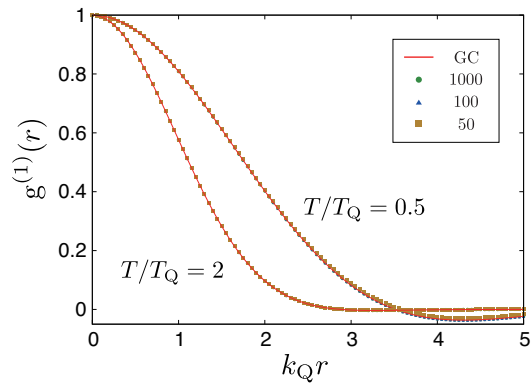}
  \caption{(Color online) One-body correlation function $g^{(1)}(r)$ of fermions with $s=1/2$ in the canonical ensemble at $T/T_{\rm Q}=0.5,2.0$ for $N=50,100,1000$ in comparison with
  those in the grand canonical ensemble (solid lines).
  \label{Fr1}}
  \end{center}
\end{figure} 
\begin{figure}[t]
\begin{center}
  \includegraphics[width=0.8\linewidth]{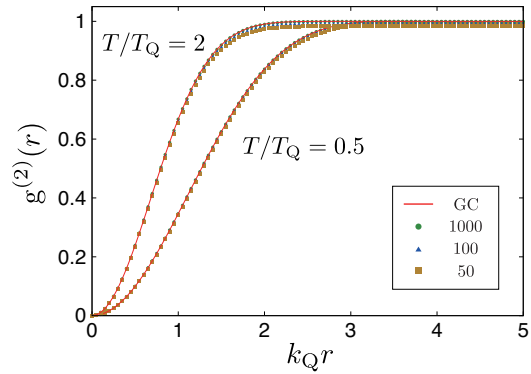}
  \caption{(Color online) Pair distribution function $g^{(2)}(r)$ of fermions with $s=1/2$  in the canonical ensemble at $T/T_{\rm Q}=0.5,2.0$ for $N=50,100,1000$ in comparison with
  those in the grand canonical ensemble (solid lines).
  \label{Fr2}}
 \end{center}
\end{figure}

We now present numerical results on $g^{(n)}(r) \equiv g^{(n)}_{\alpha\alpha}(r)$ for $n=1,2$. 
Figures \ref{Fr1} and \ref{Fr2} show the one- and two-body correlation functions of fermions with $s=1/2$, respectively,
in the canonical ensemble at $T/T_{\rm Q}=0.5,2.0$
for $N=50,100,1000$. They are compared with Eq.\ (\ref{g^(12G)}) for the grand canonical ensemble.
We observe excellent agreement between the two ensembles for fermions.
The reduction of $g^{(2)}(r)$ near the origin is due to the Pauli exclusion principle and referred to as {\it Fermi hole}, 
whose size is seen to decrease as the temperature is raised owing to the 
increase in the average kinetic energy per particle.
Tiny discrepancies between the two ensembles are seen in $g^{(2)}(r)$ for $k_{\rm Q}r\gtrsim 1.5$, which reduce in magnitude as $N$ increases.
They are responsible for the difference between the sum rules, Eqs.\ (\ref{g^(2)-sum}) and (\ref{g^(2G)-sum}).
In other words, the fluctuation term on the right-hand side of Eq.\ (\ref{g^(2G)-sum}) does not affect the short-range behavior of the 
pair distribution function of fermions.

\begin{figure}[t]
\begin{center}
  \includegraphics[width=0.8\linewidth]{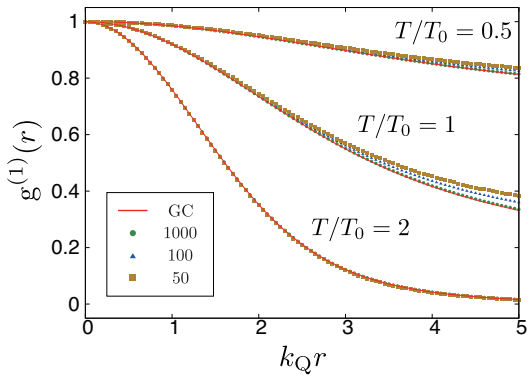}
  \caption{(Color online) One-body correlation function $g^{(1)}(r)$ of bosons with $s=0$ in the canonical ensemble at $T/T_{\rm 0}=0.5,1,2.0$ for $N=50,100,1000$ in comparison with
 those in the grand canonical ensemble (solid lines).
  \label{Br1}}
  \end{center}
\end{figure} 
\begin{figure}[t]
\begin{center}
\includegraphics[width=0.8\linewidth]{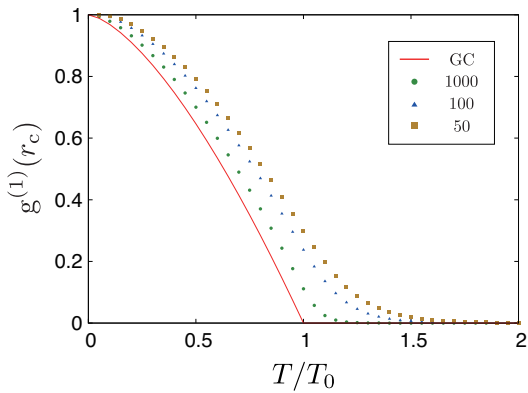}
\caption{(Color online) Temperature dependence of the one-body correlation at ${\bm r}_{\rm c}=(L/2,L/2,L/2)$ for bosons with $s=0$ in the canonical ensemble for $N=50,100,1000$
in comparison with $g^{(1{\rm G})}(\infty)$ given by Eq.\ (\ref{g^(1G)}) in the grand canonical ensemble (solid line).
\label{Fig4}}
\end{center}
\end{figure} 
\begin{figure}[t]
\begin{center}
  \includegraphics[width=0.8\linewidth]{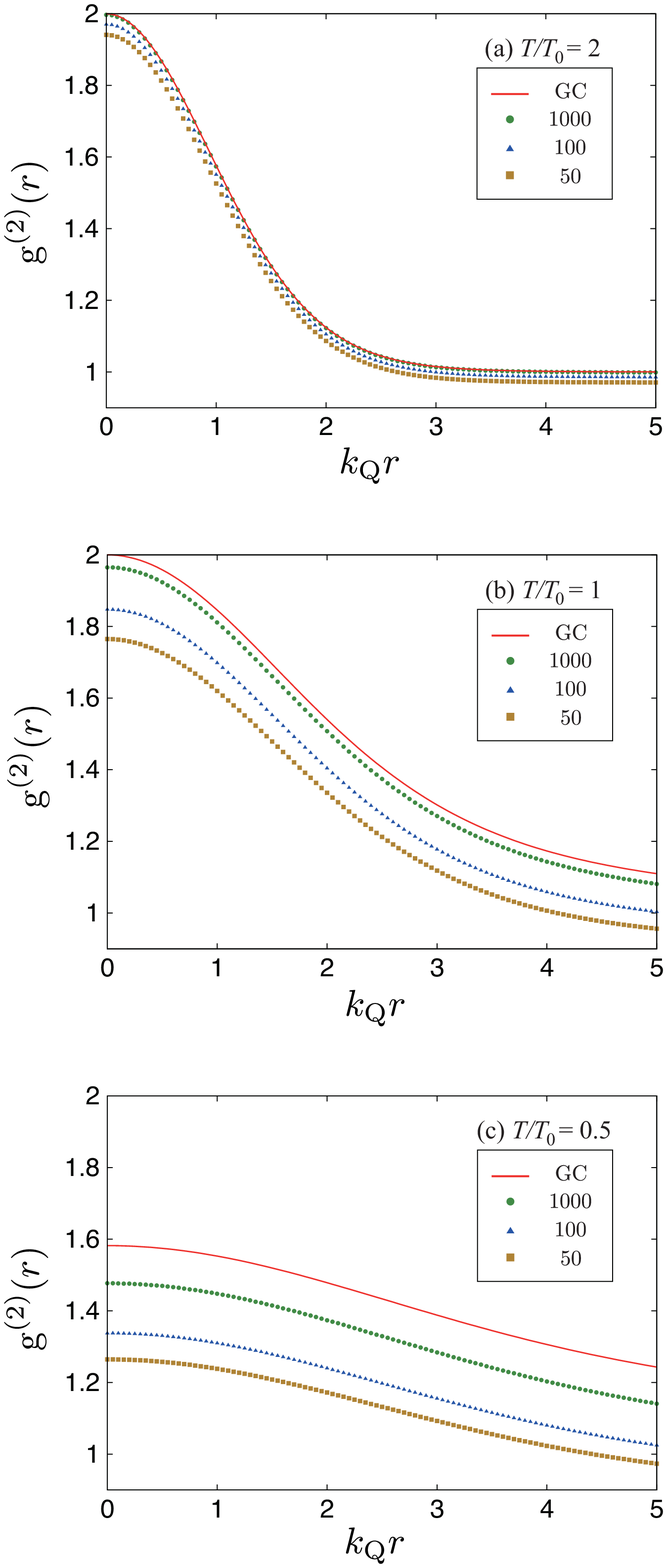}
  \caption{(Color online) Pair distribution function $g^{(2)}(r)$ of bosons in the canonical ensemble for $N=50,100,1000$ 
  at (a) $T/T_{\rm 0}=2$, (b) $T/T_{\rm 0}=1$, (c) $T/T_{\rm 0}=0.5$ in comparison with
  that in the grand canonical ensemble (solid lines).
  \label{Fig5}}
  \end{center}
\end{figure} 
\begin{figure}[t]
\begin{center}
\includegraphics[width=0.8\linewidth]{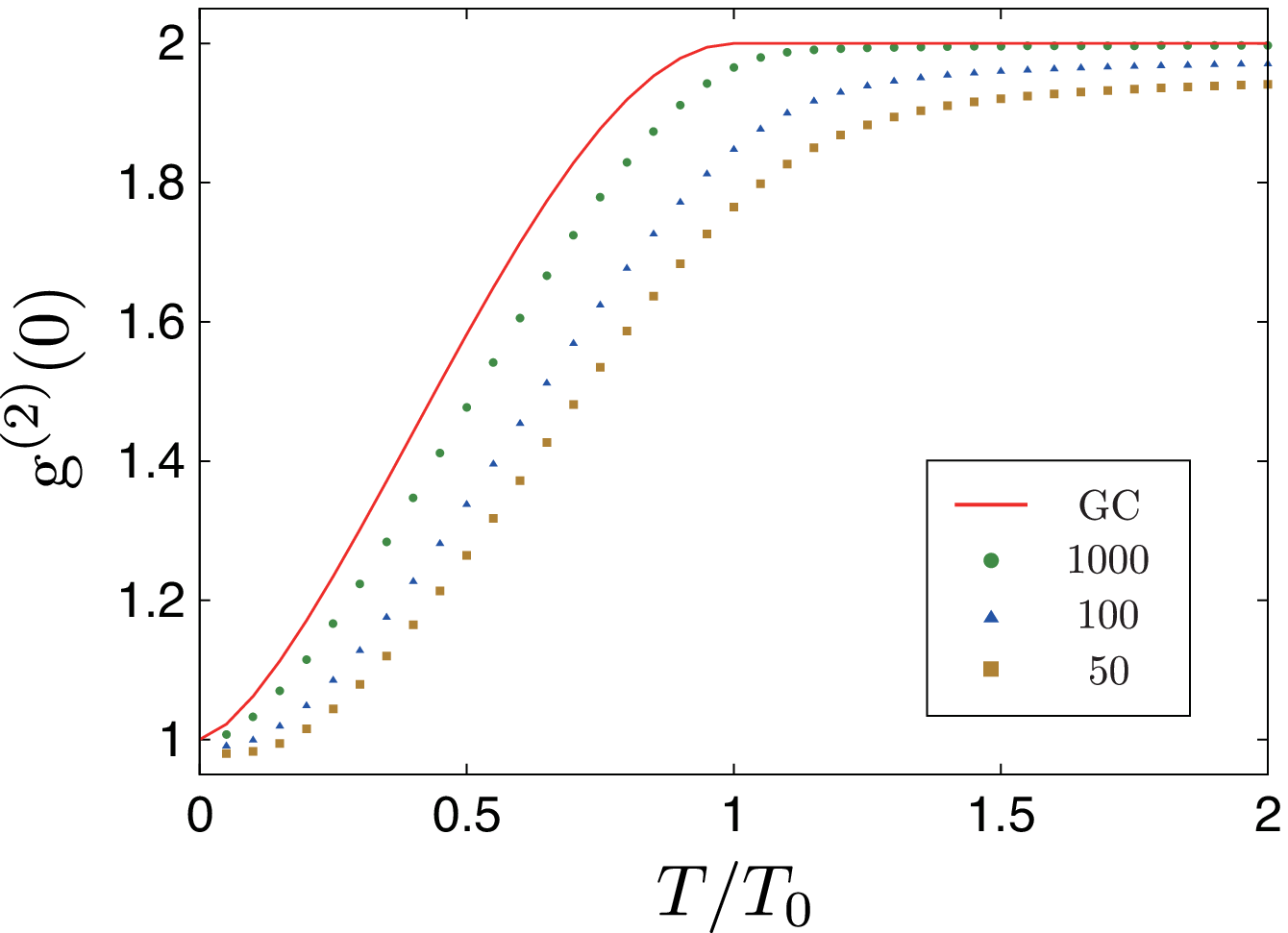}
\caption{(Color online) Temperature dependence of $g^{(2)}(0)$ for bosons in the canonical ensemble for $N=50,100,1000$ in comparison with
that in the grand canonical ensemble (solid line).
\label{Fig6}}
\end{center}
\end{figure}

We turn our attention to bosons with $s=0$. 
Figure \ref{Br1} shows the radial dependence of the one-body correlation function $g^{(1)}(r)$ at $T/T_{\rm 0}=2.0,1.0,0.5$
in the canonical ensemble for $N=50,100,1000$ in comparison with Eq.\ (\ref{g^(12G)}) for the grand canonical ensemble.
We also observe little dependence of the correlations on $N$ at $T/T_0=2.0$ without condensation,
whereas the dependence becomes stronger below $T_0$ as the condensation develops. 
To see the development of the off-diagonal long-range order monitored by $g^{(1)}(\infty)$,\cite{Penrose51,PO56,Yang} we also calculated $g^{(1)}(r)$ at ${\bm r}_{\rm c}=(L/2,L/2,L/2)$, i.e.,
at the center of the box, which is the furtherest point from the origin, as a function of temperature.
Figure \ref{Fig4} shows the temperature dependence of the one-body correlation function 
in the canonical ensemble for $N=50,100,1000$ in comparison with
$g^{(1{\rm G})}(\infty)=N_0/N$ in the grand canonical ensemble given by Eq.\ (\ref{g^(1G)}).
We observe that $g^{(1)}(r_{\rm c})$, which monitors the relative number of condensed particles, has a fairly large $N$ dependence and slowly  approaches
$g^{(1{\rm G})}(\infty)$ in the grand canonical ensemble.

Figure \ref{Fig5} shows the pair distribution function of bosons with $s=0$ at $T/T_{\rm 0}=2.0,1.0,0.5$ (from top to bottom)
in the canonical ensemble with $N=50,100,1000$.  For comparison, we also plot Eq.\ (\ref{g^(2G)}) for the grand canonical ensemble.
We observe a substantial $N$ dependence in the two-body correlation, especially for $T\ll T_0$.
Nevertheless, convergence to the grand canonical results as $N\rightarrow\infty$ are seen clearly.
The bunching of $g^{(2)}(0)=2$ above $T_0$ compared with $g^{(2)}(\infty)=1$ is due to the effective attraction
between bosons, which is seen to decrease below $T_0$. 
Figure \ref{Fig6} shows the temperature dependence of $g^{(2)}(0)$ in the canonical ensemble with $N=50,100,1000$  
in comparison with $g^{(2{\rm G})}(0)$ given by Eq.\ (\ref{g^(2G)}) for the grand canonical ensemble.
We observe steep decreases of $g^{(2)}(0)$ from $g^{(2)}(0)\approx 2$ above $T_0$ towards $g^{(2)}(0)= 1$ at $T=0$ as the condensation develops.
Moreover, the reduction starts even from above $T_0$ for smaller values of $N$.

\section{\label{sec:level4}Concluding Remarks}
We obtained an analytic expression for the reduced density matrices of quantum ideal gases in the canonical ensemble 
as Eq.\ (\ref{rho^(n)}). The formula indicates that we can also perform the Wick-decomposition-like procedure in the canonical ensemble.
Hence, it can be used to study $n$-body correlations of ideal gases  in the canonical ensemble and also to carry out perturbative calculations in the canonical ensemble.

The formula was subsequently used to clarify one- and two-body correlations of homogeneous Bose and Fermi gases.
They were also compared with those in the grand canonical ensemble, which were derived for condensed bosons by 
assuming the anomalous average $\langle\hat\psi({\bm r})\rangle\neq 0$ below the BEC temperature $T_0$. 
We confirmed that the correlations in the canonical ensemble approach those in the grand canonical ensemble
as the particle number $N$ increases towards $\infty$;
the approach is fast for fermions, whereas it becomes slower for bosons as the temperature is lowered through $T_0$.
This fact justifies the procedure of introducing the anomalous average $\langle\hat\psi({\bm r})\rangle\neq 0$ below the BEC temperature $T_0$
for bosons in the grand canonical ensemble as a mathematical means of removing the unphysical particle-number fluctuations from the grand canonical ensemble.
\\

\noindent
{\bf Acknowledgments}

K.\ T.\ is a JSPS Research Fellow, and this work was supported in part by JSPS KAKENHI Grant Number 15J01505.

\appendix

\section{Derivation of Eq.\ (\ref{Wick-GC})\label{sec:Wick-Proof}}

Let us multiply Eq.\ (\ref{rho^(n)}) by $Z_Ne^{\beta\mu N}$, perform a summation over $N=0,1,\cdots,\infty$,
and transform the right-hand side as follows:
\begin{align}
&\,\sum_{N=0}^\infty Z_N e^{\beta\mu N}\rho_N^{(n)}(\xi_1,\cdots,\xi_{n};\xi_1',\cdots,\xi_{n}')
\notag \\
=&\, \sum_{N=0}^\infty \sum_{m=n}^N e^{\beta\mu (N-m)} Z_{N-m} \sigma^{m-n}
\notag \\
&\,\times\sum_{\ell_1= 1}^m\cdots\sum_{\ell_{n}=1}^m\delta_{\ell_1+\cdots+\ell_{n},m}
\sum_{\hat P_{n}}\sigma^{P_{n}} \prod_{j=1}^n \phi_{\ell_{j}}(\xi_j,\xi_{p_j}')e^{\beta\mu\ell_j}
\notag \\
=&\, \sum_{N=0}^\infty \sum_{k=0}^{N-n} e^{\beta\mu k} Z_{k} \sigma^{N-k-n}\sum_{\ell_1= 1}^{N-k}\cdots\sum_{\ell_{n}=1}^{N-k}\delta_{\ell_1+\cdots+\ell_{n},N-k}
\notag \\
&\,\times
\sum_{\hat P_{n}}\sigma^{P_{n}}  \prod_{j=1}^n \phi_{\ell_{j}}(\xi_{p_j},\xi_{j}')e^{\beta\mu\ell_j}
\notag \\
=&\,  \sum_{k=0}^{\infty} e^{\beta\mu k} Z_{k}
\sum_{\hat P_{n}}\sigma^{P_{n}} \prod_{j=1}^n \sum_{\ell_j= 1}^\infty \sigma^{\ell_j-1} \phi_{\ell_{j}}(\xi_{p_j},\xi_j')e^{\beta\mu\ell_j}
\notag \\
=&\,  Z_{\rm G} \sum_{\hat P_{n}}\sigma^{P_{n}}\prod_{j=1}^n \sum_{q_j} \sum_{\ell_j= 1}^\infty
\sigma^{\ell_j-1} e^{-\beta(\varepsilon_{q_j}-\mu)\ell_j}\varphi_{q_j}(\xi_{p_j})
\notag \\
&\,\times \varphi_{q_j}^*(\xi_j')
\notag \\
=&\,  Z_{\rm G} \sum_{\hat P_{n}}\sigma^{P_{n}}\prod_{j=1}^n  \sum_{q_j}\frac{1}{e^{\beta(\varepsilon_{q_j}-\mu)}-\sigma}
\varphi_{q_j}(\xi_{p_j})\varphi_{q_j}^*(\xi_j')
\notag \\
=&\,  Z_{\rm G} \sum_{\hat P_{n}}\sigma^{P_{n}}\prod_{j=1}^n \langle \hat\psi^\dagger(\xi_j')\hat\psi(\xi_{p_j})\rangle.
\end{align}
The first line above is $Z_{\rm G}\rho_{\rm G}^{(n)}(\xi_1,\cdots,\xi_{n};\xi_1',\cdots,\xi_{n}')$ by definition.
We thereby obtain Eq.\ (\ref{Wick-GC}).

\end{document}